\documentclass[12pt]{article}

\usepackage{epsfig}

\usepackage{amssymb}
\usepackage{amsfonts}

\usepackage{color}
 
\def\be{\begin{equation}}
\def\ee{\end{equation}}
\def\ba{\begin{array}{c}}
\def\ea{\end{array}}

\newcommand{\kt}{\rangle}
\newcommand{\br}{\langle}

\begin{document}

\begin{center}

{\Large

Complex symmetric Hamiltonians
and exceptional points of order four and five

}

\vspace{0.8cm}

  {\bf Miloslav Znojil}

\vspace{0.2cm}

\vspace{1mm} Nuclear Physics Institute of the CAS, Hlavn\'{\i} 130,
250 68 \v{R}e\v{z}, Czech Republic

{e-mail: znojil@ujf.cas.cz}

\end{center}

\section*{Abstract}

A systematic elementary linear-algebraic construction of
non-Hermitian Hamiltonians $H=H(\gamma)$ possessing exceptional
points $\gamma=\gamma^{(EP)}$ of higher orders is proposed. The
implementation of the method leading to the EPs of orders $K=4$ and
$K=5$ is described in detail. Two distinct areas of applicability of
our user-friendly benchmark models are conjectured (1) in {\em
quantum\,} mechanics of non-Hermitian systems, or (2) in their
experimental simulations via {\em classical\,} systems (e.g.,
coupled waveguides).

\subsection*{Keywords}

non-Hermitian quantum dynamics; multilevel degeneracies; exceptional
points of higher orders; non-quantum simulations; coupled
waveguides;

\newpage

\section{Introduction}

The recent increase of popularity of Schr\"{o}dinger equations
 \be
 {\rm i}\,\partial_t |\psi\kt = H\,|\psi\kt
 \label{SEtd}
 \ee
with non-Hermitian (and, say, parameter-dependent) Hamiltonians
 \be
 H=H(\gamma) \neq H^\dagger(\gamma)
 \label{neheha}
 \ee
opened (or re-opened) a number of mathematical questions
\cite{book}. It also evoked many challenges in theoretical quantum
physics \cite{ali,Nimrod} as well as in experimental classical
physics and optics \cite{Musslimani}.

Inside the community of quantum physicists many new, promising
phenomenological models have been conjectured and studied, ranging
from their ordinary differential versions (to which the attention
has been attracted, in 1998, by the pioneering letter by Bender and
Boettcher \cite{BB}) up to their truly sophisticated quantum-field
descendants (cf., e.g., an exhaustive review \cite{Carl} of this
most ambitious project).

The Bender's and Boettcher's toy-model Hamiltonians
 $
 H_{(BB)}(\gamma) = -d^2/dx^2+V_{(BB)}(x,\gamma)
  $
were chosen non-selfadjoint but, for technical reasons, ${\cal
PT}-$symmetric,
 $
 H_{(BB)}{\cal PT}={\cal PT}H_{(BB)}
 $, with
parity ${\cal P}$ and antilinear time-reversal ${\cal T}$. Several
unexpected features of these unconventional quantum Hamiltonians
(e.g., the often-occurring reality of spectra) inspired
mathematicians who enhanced their interest in a systematic study of
similar models (cf., {\it pars pro toto}, a truly nice monograph
\cite{Trefethen}).

The idea proved inspiring also beyond quantum physics. In classical
optics, for example, the Bender's and Boettcher's ${\cal
PT}-$symmetric Schr\"{o}dinger equation was found equivalent to the
classical Maxwell equation in paraxial approximation \cite{Makris}.
Complex function $V(x)$ acquired a new physical meaning of the
complex refraction index admitting both the gain and loss of the
intensity of the beam. The quick growth of popularity of the
Maxwell-equation-related models followed. One of the reasons behind
their successful tests in the laboratory may be seen in the current
progress in nanotechnologies. This helped people to simulate various
{\it ad hoc\,} features and forms of the non-Hermiticity (e.g.,
${\cal PT}-$symmetry) {\em experimentally}, say, in the context of
physics of photonic molecules \cite{photomolec} or for the devices
composed of coupled waveguides \cite{Cart2}.

In our present paper we felt inspired by the mutual enrichment
between the quantum and non-quantum considerations, especially when
related to the concept of exceptional point (EP, cf. p.~64 in
\cite{Kato}). In section \ref{cosum} we will recall the particular
role played by the EPs in quantum physics. As an illustration we
will recall the ${\cal PT}-$symmetric version of the Bose-Hubbard
manybody model in its non-Hermitian version which has been proposed
and studied, in 2008, by Graefe et al \cite{Uwe} (cf. subsection
\ref{lecosum}). In subsequent subsection \ref{kolecosum} an
explanation will be added of the less widely known possibility of
the full theoretical compatibility between the non-Hermiticity of a
quantum Hamiltonian in an auxiliary, ``false but friendly''
\cite{SIGMA} Hilbert space (endowed with an unphysical but more
easily calculated inner product) and the unitarity of the evolution
it generates in another, amended Hilbert space using a non-trivial
Hilbert-space metric $\Theta\neq I$ in the definition of the
necessary correct and physical inner product.

In the alternative, non-quantum applications the theoretical as well
as phenomenological role of the EPs is different. In section
\ref{sedruha} we shall explain why the change of the perspective
enhances the appeal of non-Hermiticity in experimental physics as
well as in the mathematics of elementary algebraic construction
methods. In section \ref{subse}, in particular, our main message
will be then based on the turn of attention from the Bose-Hubbard
model to a more general family. Complex and symmetric
tridiagonal-matrix non-Hermitian Hamiltonians will be considered.
Using the straightforward linear-algebraic methods, a systematic
search for the EP singularities will be performed. Their exhaustive
classification will be shown fully non-numerical at $N=4$ and $N=5$.

An extensive review and analysis of the various physical aspects of
these mathematical results will finally be given in section
\ref{disskuss}, with all of this material summarized in section
\ref{sedruha7}.

\section{{\em Quantum\,} systems with exceptional points\label{cosum}}

Drawing attention to the occurrence and unfolding of exceptional
points (EPs) we intend to consider the class of finite-dimensional
matrix Hamiltonians (\ref{neheha}) with dimension $N < \infty$ and
with the special complex-symmetric structure,
 \be
 H_{mn}(\gamma)=H_{nm}^*(\gamma)\,.
 \label{cosym}
 \ee
In the context of pure mathematics such a choice could have been
found promoted by review paper \cite{Garcia} and talk \cite{Gatalk}.
From {\it loc. cit.} one deduces, i.a., an intimate relationship
between the complex symmetry of matrices and the ${\cal
PT}-$symmetry of operators.

An equally strong encouragement of our study was provided by
physicists, especially via papers \cite{Uwe} - \cite{eva} in which
the class of complex symmetric Hamiltonians has further been
narrowed to their tridiagonal-matrix subfamily such that
 \be
 H_{mn}(\gamma)=0 \ \ \ \ {\rm whenever} \ \ \ \ |m-n|\geq 2
 \,.
 \label{cosym}
 \ee
What attracted our attention to the finite-dimensional Hilbert
spaces were also certain older quantum models and results of the
physics of atomic nuclei (a recommended compact review may be found
in Ref.~\cite{Geyer}). Last but not least,  we should mention that
during our study we found a deeper physical inspiration in
Refs.~\cite{Milburn,Doppler} and in several other recent studies of
optical waveguides with gain and loss.

\subsection{${\cal
PT}-$symmetric Bose-Hubbard model of Graefe et
al~\cite{Uwe}\label{lecosum}}

In phenomenological context the proposal and study of the ${\cal
PT}-$symmetric Bose-Hubbard model of Ref.~\cite{Uwe} was motivated
by the search for an elementary simulation of the process of the
Bose-Einstein condensation. Up to the purely numerically tractable
interaction term the non-Hermitian Hamiltonian of the model was
chosen in the form
 \be
 H_{(BH)}(\gamma)=
  2\,(-{\rm i}\gamma \,L_z + L_x) 
  \label{Uwemo}
 \ee
of the complex linear combination of the two angular-momentum
generators $L_{z,x}$ of the real Lie algebra $su(2)$. The underlying
representation theory enables one to treat operator (\ref{Uwemo}) as
decomposed into an infinite family of finite-dimensional $N$ by $N$
matrices
 \be
 H^{(2)}_{(BH)}(\gamma)=
 \left[ \begin {array}{cc} -i{\it \gamma}&1
 \\\noalign{\medskip}1&i{\it
 \gamma}
 \end {array} \right]\,,
 \label{dopp2}
 \ee
 \be
H^{(3)}_{(BH)}(\gamma)=\left[ \begin {array}{ccc} -2\,i\gamma&
\sqrt{2}&0\\\noalign{\medskip}\sqrt{2}&0&
\sqrt{2}\\\noalign{\medskip}0&\sqrt{2}&2\,i\gamma\end
{array} \right]\,,
  \label{3wg}
 \ee
etc. An inessential change of parameter $\gamma \to \sqrt {z} $ can
be recommended in the subsequent items, i.e., in
 \be
 H^{(4)}_{(BH)}(z)= \left[ \begin {array}{cccc} -3\,i\sqrt {z}&\sqrt
 {B}&0&0
 \\\noalign{\medskip}\sqrt {B}&-i\sqrt {z}&\sqrt {A}&0
 \\\noalign{\medskip}0&\sqrt {A}&i\sqrt {z}&\sqrt {B}
 \\\noalign{\medskip}0&0&\sqrt {B}&3\,i\sqrt {z}\end {array}
 \right]\,
 \label{tripa}
 \ee
with $B=3$ and $A=4$ as well as in
 \be
 H^{(5)}_{(BH)}(z)=\left[
 \begin {array}{ccccc} -4\,i\sqrt {z}&\sqrt {B}&0&0&0
 \\\noalign{\medskip}\sqrt {B}&-2\,i\sqrt {z}&\sqrt {A}&0&0
\\\noalign{\medskip}0&\sqrt {A}&0&\sqrt {A}&0\\\noalign{\medskip}0&0&
\sqrt {A}&2\,i\sqrt {z}&\sqrt {B}\\\noalign{\medskip}0&0&0&\sqrt
{B}&4 \,i\sqrt {z}\end {array} \right]\,
 \label{petpa}
 \ee
with $B=4$ and $A=6$, etc. Such a change of notation is motivated by
the simplification of the respective secular equations (see below).
Still, the possibility of the alternative choice of sign of $\gamma
\to -\sqrt {z} $ should be kept in mind as a helpful symmetry of the
Bose-Hubbard model (see also picture Nr. 1 in \cite{Uwe}). Due to
this symmetry and due to the reality of the spectra, all of the
matrices (\ref{dopp2}) - (\ref{petpa}) (etc) may be declared
eligible, together with their $\gamma \to -\gamma\,$ counterparts,
as non-numerical toy-model generators of quantum evolution.

\subsection{The procedure of Hermitization\label{kolecosum}}

In the conventional studies of quantum evolution one usually
requires that it is unitary. For our sample generators
$H^{(N)}_{(BH)}(\gamma)$ (i.e., for the diagonalizable, hiddenly
Hermitian Bose-Hubbard Hamiltonians) this means that the spectrum
must be real. The authors of Ref.~\cite{Uwe} emphasized that the
latter condition is satisfied if and only if $-1 < \gamma < 1$. For
the conventional non-negative $\gamma\geq 0$ and real $z \geq 0$
they concluded that every element $ H^{(N)}_{(BH)}(z)$ of the
sequence is exactly solvable and, moreover, that each toy-model
Hamiltonian $H^{(N)}_{(BH)}(z)$ in the family possesses a unique and
real {\em exceptional point\,} $z^{(EP)}=1$ of order $N$.


For the admissible parameters $\gamma \in (-1, 1)$ one can find a
(non-unique) Hermitian and positive matrix of Hilbert-space metric
$\Theta$ such that
 \be
 H^\dagger \,\Theta = \Theta \,H\,
 \label{requi}
 \ee
i.e., such that $H=H^{(N)}_{(BH)}(\gamma)$ may be declared Hermitian
with respect to an amended inner product $\br
\cdot|\Theta\,\cdot\kt$. The details of the theory using nontrivial
metrics $\Theta \neq I$ may be found, say, in reviews
\cite{ali,Geyer}. For our present purposes it is only necessary to
emphasize that the first principles of the theory remain unchanged.
Thus, any observable phenomenon must still be represented by a
hiddenly Hermitian operator  $\Lambda$ such that $\Lambda^\dagger
\,\Theta = \Theta \,\Lambda$. In other words, only the initial
knowledge of the Hamiltonian-dependent metric $\Theta=\Theta(H)$
makes the model theoretically complete.

Needless to add, the necessary guarantee of validity of
Eq.~(\ref{requi}) is far from trivial in practice. The difficulty of
the construction of $\Theta=\Theta(H)$ is, incidentally, one of the
reasons why the matrix phenomenological models with small dimensions
$N \ll \infty$ are so important, both in the theory and in its
applications.

\subsubsection{The construction of metric at $N=2$}

At $N=2$ we may insert quantum Hamiltonian $H^{(2)}_{(BH)}(\gamma)$
of Eq.~(\ref{dopp2}) in Eq.~(\ref{requi}). The routine solution of
this linear algebraic problem yields {\em all\,} of the eligible
(i.e., positive definite) metrics. Up to an overall inessential
multiplicative factor they are all defined by the following
one-parametric formula
 \be
 \Theta^{(2)}(\beta)= I^{(2)} +\left[ \begin {array}{cc} 0&\beta+i\,\gamma
 \\\noalign{\medskip}\beta-i\,\gamma&0
 \end {array} \right]\,,\ \ \ \ \ -\sqrt{1-\gamma^2}<\beta< \sqrt{1-\gamma^2}\,.
 \label{medopp2}
 \ee
Obviously, in the arbitrarily small vicinities of
$\gamma=\gamma^{(EP)}$  such a metric only exists, in the spirit of
Ref.~\cite{lotor}, after the minimally anisotropic metric constant
choice of $\beta=0$.

\subsubsection{The construction of metric at $N=3$}

The fully analogous treatment of the next quantum system with
Hamiltonian (\ref{3wg}) (in which we abbreviate $g=\sqrt{2}\gamma$)
will lead to the two-parametric family of metrics
 \be
 \Theta^{(3)}(\beta,\delta)= I^{(3)} +\left[ \begin {array}{ccc}
  0&\beta+i\,g&
 \delta+i\,g\,\beta\\
 \noalign{\medskip}\beta-i\,g&\delta+{g}^{2}&\beta+i\,g\\
 \noalign{\medskip}\delta-i\,g\,\beta&\beta-i\,g&0\end {array}
 \right]\,.
  \label{me3wg}
 \ee
In the spirit of Ref.~\cite{lotor} one may again prefer a
``minimally anisotropic'' choice of $\beta=0$ and $\delta=0$ which
yields the metric with eigenvalues
 \be
 \theta_0=1\,,
 \ \ \theta_\pm =
 1+\frac{1}{2}\,{g}^{2}\pm \frac{1}{2}\,\sqrt
 {8\,{g}^{2}+{g}^{4}}\,.
 \ee
Obviously, the resulting special metric $\Theta^{(3)}(0,0)$ still
exists up to the maximal admissible (i.e., real-energy-admitting)
non-Hermiticity limit of $g \to g^{(EP)}=1$.

An entirely analogous procedure will also work at any higher matrix
dimension $N$.

\section{Exceptional points in {\em non-quantum\,} optics\label{sedruha}}

Schr\"{o}dinger-like equations and their solutions appear in many
non-quantum branches of physics. Thus, in the physical context of
classical optics the quantum effects may be simulated via microwave
devices \cite{micro} or coupled waveguides \cite{wg}. Once these
devices prove characterized by a symmetry between the gain and loss
in the medium, the mathematics of solutions becomes shared with the
quantum theories exhibiting the combined parity times time reversal
symmetry {\it alias\,} ${\cal PT}-$symmetry \cite{Carl}. Thus, after
one leaves the unitary quantum mechanics and after one turns
attention to classical optics, the related
Schr\"{o}dinger-to-Maxwell change of the meaning of the
Schr\"{o}dinger-like evolution Eq.~(\ref{SEtd}) gets accompanied by
the slightly easier, formally less restrictive mathematics. For
example, the ``wave function'' solutions $|\psi\kt$ need not be
normalizable anymore.

In the new context it is crucial that the evolution generated by the
Hamiltonian-resembling operators $H=H^{(N)}_{}(\gamma)$ need not be
required unitary. The spectra of ``energies'' may be complex while
the components of the ``wave functions'' themselves may become
directly measurable \cite{Musslimani}. Last but not least, the
simplification of mathematics may be accompanied by the feasibility
of experimental setups in which the generators are allowed
time-dependent, $H=H(t)$ \cite{timedep}.

\subsection{Two coupled waveguides}

The theoretical studies of the non-stationary systems as well as the
realizations of the experiments remain highly nontrivial.
Fortunately, this direction of research also leads to multiple
surprising results reconnecting the quantum and classical physics.
For illustration one may recall the really surprising recent
discovery of the failure of applicability of the intuitive adiabatic
hypothesis in the non-Hermitian setting \cite{Milburn} which was
first reported during the 15th International Workshop on
Pseudo-Hermitian Hamiltonians in Quantum Physics in Palermo
\cite{Palermo} in 2015.

For explanation, just a version of the two-by-two toy model of
Eq.~(\ref{dopp2}) proved sufficient. In the recent compact review
\cite{Doppler} the authors described an interesting application of
the new phenomenon to the description of scattering of the classical
(say, electromagnetic) waves through a two-mode waveguide device. A
comparatively satisfactory theoretical explanation of the phenomenon
of the breakdown of adiabatic approximation has been achieved via
the numerical solution of the manifestly time-dependent $N=2$
evolution rule (\ref{SEtd}).

On this background a number of ``elusive effects'' has been
predicted, resulting from the fact that the Hamiltonian matrix
itself is symmetric but not real. Indeed, the requirement of its
reality would make it Hermitian so that all of the ``elusive
effects'' would disappear. Theoretically the device was described by
a slightly modified version
 \be
 H(\delta,g,\gamma_1,\gamma_2)=
 \left[ \begin {array}{cc}\delta -i{\it \gamma_1}/2&g
 \\\noalign{\medskip}g&-i{\it
 \gamma_2}/2
 \end {array} \right]\,
 \label{dopp}
 \ee
of the most elementary model (\ref{dopp2}). All of the four
parameters (viz., $\delta$ responsible for the so called detuning,
$g$ denoting the mutual symmetric coupling of the modes, and
$\gamma_{1}$ and $\gamma_{2}$ gauging the losses in the medium of
the two respective waveguides) were chosen real.

The model yields a pair of instantaneous eigenenergies
$E_\pm(z)=E_\pm[\delta(z),g(z),\gamma_1(z),\gamma_2(z)]$ which are
distinct and complex in general. The secular equation is trivial
yielding the two values $E_\pm(z)$ in closed form.
These values may be perceived
as evaluations of a two-sheeted
analytic function $\mathbb{F}(z)$. According to Kato \cite{Kato}
such an explicit representation of the spectrum also enables us
to localize the so called ``exceptional points''
$z^{(EP)}$ at which the two levels of the
energy happen to merge,
$E_{+}(z^{(EP)})=E_{-}(z^{(EP)})$.

In \cite{Doppler}, the authors performed the search. In their
two-by-two matrix model (\ref{dopp}) the analytic-function
representation of the energies appeared to have the square-root form
near $z^{(EP)}$, with $\mathbb{F}(z) \sim
\sqrt{z-z^{(EP)}}$. In an experimental setup the ``no-detuning'' choice of
$\delta=0$
and the $z-$independent choice of the
losses $\gamma_{1,2}$ enabled the authors to determine
the degeneracy-responsible EP couplings in closed form,
 \be
 g=g[z^{(EP)}]= \frac{1}{4}
 \left (
 \gamma_1-\gamma_2
 \right )\,.
 \ee
The existence of the
non-diagonalizable EP limit of the Hamiltonian
 \be
 \lim_{z \to z^{(EP)}}H[\delta(z),g(z),\gamma_1(z),\gamma_2(z)]
 =\frac{1}{4}
 \left[ \begin {array}{cc} -2\,i{\it \gamma_1}&{\it \gamma_1}-{\it
 \gamma_2}\\\noalign{\medskip}{\it \gamma_1}-{\it \gamma_2}&-2\,i{\it
 \gamma_2}
 \end {array} \right]\,
 \label{doppep}
 \ee
has been also found detectable, experimentally, due to the
mathematical property of the eigenvalues which form, in the EP
vicinity, the so called cycle of period two (cf. p.~65 in
\cite{Kato}). In the context of physics the latter mathematical
peculiarity of the model enabled the authors to reveal the limits of
the applicability of the conventional adiabatic hypothesis in the
non-Hermitian and non-stationary dynamical setting~\cite{Doppler}.

Multiple analogous searches for the signatures of the two-sheeted
nature of the energy Riemann surfaces were performed, recently, by
many independent experimental groups \cite{experdva}. In the
language of mathematics the existence of the EP2 singularity means
that one might perform such a variation of the parameters that the
system would be forced to circumscribe its EP2 branch point and to
move to the second sheet of the Riemann surface. In order to force
the wave function to return to its initial value, one has to
circumscribe the EP2 singularity twice.

In an alternative experimental setup one could try to force the
system to pass strictly through the EP2 singularity. This would lead
to the coalescence $|\psi_{0}(z^{(EP)}\kt=|\psi_{1}(z^{(EP)})\kt$ of
wave functions, tractable as a simulation of a quantum phase
transition \cite{Denis}. Beyond quantum world, the effect is equally
interesting. In the context of waveguides, for example, one could
even force the classical photons to stop at EP2 \cite{light}.

An analogous theoretical as well as experimental analysis would
become much less easily accessible in the general dynamical
scenarios characterized by the Kato's EPs of order $K > 2$. Any
experimental study of the dynamical $K>2$ mode switching would
require a rather sophisticated equipments working, say, with
complicated but still tractable waveguide systems as sampled, say,
by Fig. Nr. 6 in Ref.~\cite{Teimo}.

%
%
%
%

%

\subsection{Three coupled waveguides \label{drususedruha}}

In paper \cite{Cart} the authors contemplated an experimentally
feasible arrangement of a coupled {\em triplet\,} of semiconductor
waveguides. Their system (exhibiting, in addition, the so called
${\cal PT}-$symmetric distribution of the gain and loss in the
medium of the waveguides) found a formal theoretical description in
another elementary schematic Hamiltonian, viz., matrix (\ref{3wg}).
This is a complex and symmetric matrix, not too dissimilar to its
two-dimensional predecessor (\ref{dopp2}). In Ref.~\cite{Cart} we
may read that ``In principle the approach of extending the system
with additional waveguides \ldots can be continued [but] \ldots all
further extensions should first be studied in [simplified]
approaches.''

These sentences may be re-read as the most concise formulation of
the aims of our present paper. One still has to expect that any
extension of the EP2-related results to the case of the $K-$sheeted
Riemann-surface scenarios with $K>2$ will be complicated. Even in
the case of $K=3$ the proper design of the experimental setup
required a careful fine-tuning of parameters \cite{eva,studythree}.
On positive side, a strong encouragement comes from the observation
that a decisive theoretical progress has been achieved after the
scope of the experiment-oriented searches had been restricted to the
tridiagonal complex symmetric models as sampled by Eq.~(\ref{3wg})
above. In this sense, a decisive theoretical step forward has been
made by the authors of Ref.~\cite{eva}. Subsequently, the practical
experimental appeal of the tridiagonal matrix model has been
emphasized in \cite{Cart2} and \cite{Cart}.

All of the latter $K=3$ projects were aimed at a slow motion along a
path over the three-sheeted Riemann surface while circumscribing,
three times, the carefully localized exceptional point of order
three (EP3). The details may be found in Figure Nr. 3 of
Ref.~\cite{Cart}. What the latter studies revealed was, first of
all, the phenomenon of the characteristic permutation of the
components of $|\psi\kt$. This offered a signature of the
coalescence of all of the three eigenfunctions at $z=z^{(EP3)}$. It
also appeared to make sense to change the parameters in
 $$
 H=H^{(3)}(z)=\left[ \begin {array}{ccc} -i\sqrt {z}&\sqrt
 {A(z)}&0\\\noalign{\medskip}\sqrt {A(z)}&0&\sqrt
 {A(z)}\\\noalign{\medskip}0&\sqrt { A(z)}&i\sqrt {z}\end {array}
 \right]\,.
 $$
This simplified the secular polynomial as well as its three energy
roots $E_\pm(z) = \pm \sqrt {2\,A(z)-z}$ and  $ E_0= 0 \neq E_0(z)$.
One could spot the EP3 singularity at $z=z^{(EP3)}=1$ and
$A(1)=1/2$,
 \be
 \lim_{z \to z^{(EP3)}}H^{(3)}(z)
 =H_{(EP3)}^{(3)}=
 \left[ \begin {array}{ccc}
 -i&1/\sqrt {2}&0
 \\\noalign{\medskip}1/\sqrt {2}&0&1/\sqrt {2}
 \\\noalign{\medskip}0&1/\sqrt {2}&i
 \end {array} \right]\,.
 \label{hyopp}
 \ee
An extensive analysis of consequences may be found in
Ref.~\cite{eva}. It made us to conclude that the models using $K>3$
deserve to be studied along similar lines.

%
%

\section{Tridiagonal complex symmetric Hamiltonians revisited\label{subse}}

Theoretical papers \cite{Uwe,admis} paid attention to the systems
which are assigned a multi-sheeted Riemann surface $\mathbb{F}$
admitting the existence and, perhaps, experimental detection of EPs
of the $K-$th order. At any integer $K$ one can work, locally, with
the $K-$sheeted analytic function $\mathbb{F}(z)\sim
(z-z^{(EP)})^{1/K}$ representing the energy eigenvalues. In what
follows we will try to support this point of view constructively.

\subsection{Four by four Hamiltonian matrices}

Let us pick up $K=4$ and contemplate the eligible Hamiltonian
matrices in the tridiagonal and complex symmetric form of
Eq.~(\ref{tripa}) above. The key merit of this choice is that it
yields energies $E = \pm \sqrt{x}$ determined in terms of roots of
the exactly solvable secular equation
 \be
 {x}^{2}+10\,xz-2\,Bx-Ax+9\,{z}^{2}+6\,Bz-9\,zA+{B}^{2}=0\,.
 \label{secue4}
 \ee
The availability of closed formula
 \be
 x_\pm =B-5\,z+\frac{1}{2}\,A\pm \frac{1}{2}
 \,\sqrt {-64\,Bz+4\,BA+64\,{z}^{2}+16\,zA+{A}^{2}}\,
 \label{formux}
 \ee
reduces the search for the EP4 confluence of the roots $x_\pm
(z^{(EP4)})=0$ to the analysis of Eq.~(\ref{formux}), yielding the
following two algebraic equations for three unknowns,
 \be
 B-5\,z+1/2\,A=0\,,
  \ \ \ \
  -64\,Bz+4\,BA+64\,{z}^{2}+16\,zA+{A}^{2}=0\,.
  \label{dverovnice}
 \ee
They have the two independent EP4 solutions, viz., the well known
Bose-Hubbard solution of Ref.~\cite{Uwe},
 \be
 B^{(EP4)} = 3\,z\,, \ \ \ \  A^{(EP4)} = 4\,z
 \label{prve}
 \ee
and the new solution
 \be
 B^{(EP4)} = -27\,z\,, \ \ \ \ A^{(EP4)} = 64\,z\,.
 \label{druhe}
 \ee
As long as both of these solutions are non-numerical, they may
easily be analyzed in detail.

\subsubsection{Bose-Hubbard model revisited}


Once we pick up the first solution and set, tentatively, $B = 3$ and
$A = 4$ we arrive at the one-parametric family of Hamiltonians
 \be
 H=H^{(4)}(z)=\left[
 \begin {array}{cccc} -3\,i{\it \sqrt{z}}&\sqrt {3}&0&0
 \\\noalign{\medskip}\sqrt {3}&-i{\it \sqrt{z}}&2&0\\
 \noalign{\medskip}0&2&i{
 \it \sqrt{z}}&\sqrt {3}\\
 \noalign{\medskip}0&0&\sqrt {3}&3\,i{\it \sqrt{z}}
 \end {array} \right]
 \label{naseeq}
 \ee
with energies
 $$
 E_{\pm,\pm}(z)= \pm (2 \pm 1) \sqrt{1-{z}}\,.
 $$
All four of them remain real for ${z} \in (0,1)$, and all of them
vanish at ${z}=z^{(EP4)}= 1$. Beyond this point, all of the levels
become purely imaginary so that the model is to be assigned the
physical interpretation of an open quantum system or of a device in
classical optics.

Inside the interval of ${z} \in (0,1)$, on the contrary, the
guaranteed reality of the spectrum leads to the possibility of the
construction of the metric which would render possible the
consistent unitary-evolution interpretation of the system in quantum
setting. The authors of study \cite{Uwe} left, unfortunately, this
construction of $\Theta(H)$ as well as the discussion of its
properties to the readers as an elementary exercise. Here, we shall
return to this point in paragraph \ref{papapa} below.

\subsubsection{Jordan blocks}

By definition, the Bose-Hubbard Hamiltonian (\ref{naseeq}) ceases to
be diagonalizable at the EP4 singularity. It can only be assigned
there the canonical four-dimensional non-diagonal Jordan-block
representation. In general, at any EP energy degeneracy of order $K$
we may postulate
 \be
 H^{(K)}(z^{(EPK)}) Q^{(K)} = Q^{(K)} J^{(K)}(E)\,.
 \label{realt}
 \ee
The symbol $J^{(K)}(E)$ stands here for the $K$ by $K$ Jordan block
 \be
 J^{(K)}(E)=\left [\begin {array}{ccccc}
    E&1&0&\ldots&0
 \\{}0&E&1&\ddots&\vdots
 \\{}0&0&E&\ddots&0
 \\{}\vdots&\ddots&\ddots&\ddots&1
 \\{}0&\ldots&0&0&E
 \end {array}\right ]\,.
 \label{hisset}
 \ee
The other symbol $Q=Q^{(K)}$ is Hamiltonian-dependent. It denotes
the object called transition matrix. For example, in our present
Bose-Hubbard illustrative example with $E=E^{(EP4)}=0$ and
$z^{(EP4)}=1$ it is entirely routine to evaluate
 $$
 Q^{(4)}=\left[ \begin {array}{cccc} 6\,i{z}^{3/2}&-6\,z&-3\,i\sqrt
 {z}&1
 \\\noalign{\medskip}-6\,{z}^{3/2}\sqrt {3}&-4\,iz\sqrt {3}&\sqrt {3}
 \sqrt {z}&0\\\noalign{\medskip}-3\,i{z}^{3/2}\sqrt {3}\sqrt
 {4}&\sqrt {3}z\sqrt {4}&0&0\\\noalign{\medskip}3\,{z}^{3/2}\sqrt
 {4}&0&0&0
 \end {array} \right]\,
 $$
Such a transition matrix plays a key role in the
perturbation-expansion analysis of Schr\"{o}dinger equations
(\ref{SEtd}) in the vicinity of EPs of any order $K\geq 2$. The
physical consequences as well as the mathematical details may be
found explained in Ref.~\cite{Uwe} and, beyond the Bose-Hubbard
illustrative example, in  Ref.~\cite{admis}.

\subsubsection{Generalized Bose-Hubbard model}

\begin{figure}[h]                     
\begin{center}                         
\epsfig{file=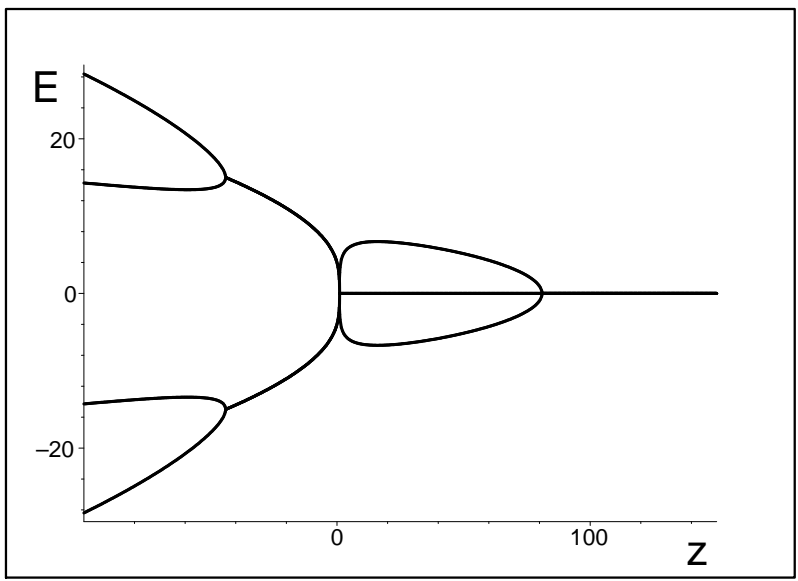,angle=270,width=0.6\textwidth}
\end{center}                         
\caption {Real parts of eigenvalues $E_n(z)$ of the non-Hermitian
$N=4$ matrix $H^{(N)}({z})$ of Eq.~(\ref{26}).
 \label{ee1}}
\end{figure}
\begin{figure}[h]                     
\begin{center}                       

\epsfig{file=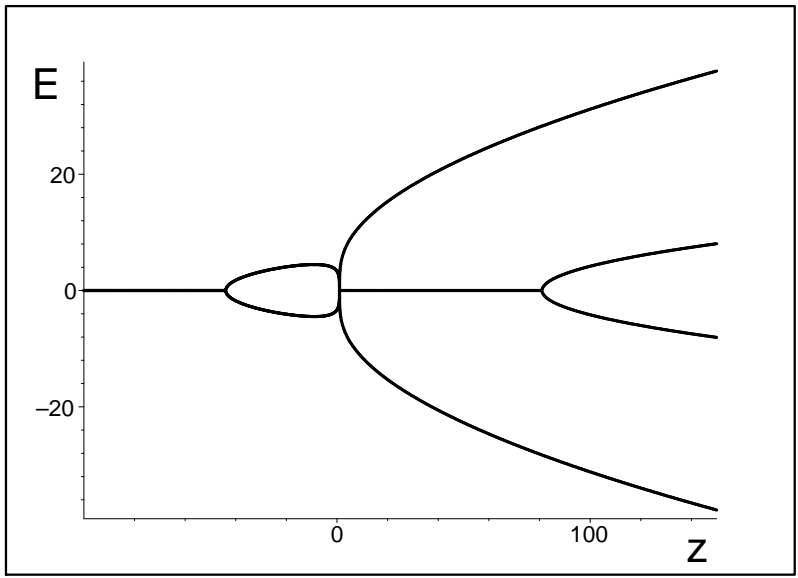,angle=270,width=0.6\textwidth}

\end{center}                         
\caption {Imaginary parts of eigenvalues $E_n(z)$ of the
non-Hermitian $N=4$ matrix $H^{(N)}({z})$ of Eq.~(\ref{26}).
 \label{ee2}}
\end{figure}

Our first new, non-Bose-Hubbard result is that we may insert $B =
-27$ and $A = 64$ in Eq.(\ref{tripa}). This yields the
EP4-supporting complex symmetric toy model
 \be
 H^{(4)}({z})=\left[
 \begin {array}{cccc} -3\,i{\it \sqrt{z}}&3\,i\sqrt {3}&0&0
 \\\noalign{\medskip}3\,i\sqrt {3}&-i{\it \sqrt{z}}&8&0
 \\\noalign{\medskip}0&
 8&i{\it \sqrt{z}}&3\,i\sqrt {3}\\\noalign{\medskip}0&0&3\,i\sqrt
 {3}&3\,i{ \it \sqrt{z}}\end {array} \right]
 \label{26}
 \ee
with energies
 \be
 E_{\pm,\pm}(z)= \pm \sqrt {5-5\,{{\it {z}}}\pm 4\,\sqrt {-44+43\,
 {{\it {z}}}+{{\it {z}}} ^{2}}}\,.
 \label{drua}
 \ee
The analysis is again straightforward. Graphically, the
$z-$dependence of the real and imaginary parts of the eigenvalues of
matrix (\ref{26}) is displayed in respective Figs.~\ref{ee1} and
~\ref{ee2}. The picture shows that all the four energies vanish at
${z}=1$ (EP4). The inner square root expression remains purely
imaginary at smaller positive ${z} \in (0,1)$ so that all of the
four related energies are complex. Otherwise, the inner square root
expression is real so that one only has to distinguish between the
interval of ${z} \in (1,81)$ (in which we have two real and two
purely imaginary energies) and the interval of ${z} \in
(81,\infty)$.


%

%
%
%

%


In the latter interval all of the four energy roots are purely
imaginary so that an {\it ad hoc\,} premultiplication of matrix
(\ref{26}) by imaginary unit would make the spectrum real, rendering
the resulting complex symmetric matrix $\tilde{H}^{(4)}({z})={\rm
i}H^{(4)}({z})$ eligible as a correct and physical hiddenly
Hermitian generator of unitary evolution in a new, slightly exotic
quantum model with just two imaginary matrix elements.

One can summarize that at positive $z\geq 0$ the model exhibits the
EP4 degeneracy at ${z}=1$ and the EP2 degeneracy at ${z}=81$. It is
worth adding that at $z=1$ the model satisfies Eq.~(\ref{realt})
with the following transition matrix
 $$
 Q^{(4)}=\left[
 \begin {array}{cccc} 216\,i{z}^{3/2}&-36\,z&-3\,i\sqrt {z}&1
 \\\noalign{\medskip}72\,z\sqrt {-3\,z}&-12\,i\sqrt {z}\sqrt {-3\,z}&3
 \,\sqrt {-3\,z}&0\\\noalign{\medskip}-9\,iz\sqrt {-3\,z}\sqrt
 {64}&3\, \sqrt {-3\,z}\sqrt {64}\sqrt
 {z}&0&0\\\noalign{\medskip}-27\,{z}^{3/2} \sqrt {64}&0&0&0\end
 {array} \right]\,.
 $$
In the case of the hiddenly Hermitian Hamiltonian
$\tilde{H}^{(4)}({z})$ only two of the energies vanish in the limit
$z \to z^{(EP2)} =81$ so that the perturbation study of its vicinity
would only be based on the use of the transition matrix of rank two.

\subsection{Five by five Hamiltonian matrices}

%

With $N=5$ and with the Bose-Hubbard choice of $B = 4$, $A = 6$ and
$z\geq 0$ in (\ref{petpa}) we have to deal with the tridiagonal
complex symmetric Hamiltonian
possessing, up to a constant level $E_0=0$, the following four
$z-$dependent energy eigenvalues
 $$
 E_{\pm,\pm}(z)= \pm (3 \pm 1) \sqrt{(1-z)}\,.
 $$
They stay real for $z \in (0,1)$, and all of them vanish at $z=1$.
The EP5 presence is readily verified. The $z-$dependent quadruplet
becomes purely imaginary at any $z>1$ \cite{Uwe}. At $z =
z^{(EP5)}=1$ the Hamiltonian ceases to be diagonalizable. With
transition matrix
 $$
 Q^{(5)}=\left[ \begin {array}{ccccc} 24&24\,i&-12&-4\,i&1
 \\\noalign{\medskip}48\,i&-36&-12\,i&2&0
 \\\noalign{\medskip}-24\,\sqrt {6}&-12\,i\sqrt {6}
&2\,\sqrt {6}&0&0\\\noalign{\medskip}-48\,i&12&0&0&0
\\\noalign{\medskip}24&0&0&0&0\end {array} \right]\,
 $$
it may still be shown to satisfy Eq.~(\ref{realt}).

\subsubsection{Bose-Hubbard model revisited}

Even if we do not specify parameters $A$ and $B$, the five-by-five
complex symmetric tridiagonal matrix Hamiltonian (\ref{petpa})
preserves the symmetry of the energies $E = \pm \sqrt{x}$. Keeping
in mind the $z-$independent root $x_0=0$, i.e., $E_0=0$ we have
still an utterly elementary secular equation
 \be
 {x}^{2}+20\,xz-2\,xB-2\,Ax
 +2\,AB-32\,zA+64\,{z}^{2}+16\,zB+{B}^{2}=0\,.
 \label{secue5}
 \ee
It is straightforward to deduce that
 $$
 x_\pm =B-10\,z+A \pm \sqrt {-36\,zB+36\,{z}^{2}+12\,zA+{A}^{2}}\,.
 $$
The search for the EP5 confluence
%
proceeds again via the pair of relations
 \be
 -36\,zB+36\,{z}^{2}+12\,zA+{A}^{2}=0\,,
 \ \ \ \ \
 B-10\,z+A=0
  \label{dverovnice5}
 \ee
with solutions
 \be
 {B^{(EP5)} = 4\,z\,, \ \ \ A^{(EP5)} = 6\,z}
 \label{prve5}
 \ee
and
 \be
  {A^{(EP5)} = -54\,z\,, \ \ \ B^{(EP5)} = 64\,z}\,.
  \label{druhe5}
 \ee
This means that the conclusions may be expected to remain
qualitatively the same as at $N=4$ above.

\subsubsection{Generalized Bose-Hubbard model}



The non-Bose-Hubbard choice of $B = 64$ and $A = -54$ yields the
``anomalous'' Hamiltonian
 $$
 H^{(5)}(z)=\left[ \begin {array}{ccccc} -4\,i\sqrt {z}&8&0&0&0
 \\\noalign{\medskip}8&-2\,i\sqrt {z}&3\,i\sqrt {6}&0&0
\\\noalign{\medskip}0&3\,i\sqrt {6}&0&3\,i\sqrt {6}&0
\\\noalign{\medskip}0&0&3\,i\sqrt {6}&2\,i\sqrt {z}&8
\\\noalign{\medskip}0&0&0&8&4\,i\sqrt {z}\end {array} \right]\,.
 $$
Up to the constant ``observer energy'' $E=0$, the quadruplet of its
nontrivial energies reads
 $$
 E_{\pm,\pm}(z)= \pm \sqrt {-10\,z+10 \pm 6\,\sqrt
 {{z}^{2}-82\,z+81}}\,.
 $$
They vanish at $z= 1$ (EP5). One can detect the occurrence of two
coupled EP2 singularities at $z=81$, with the two non-vanishing,
purely imaginary energies $E^{(EP2)} _\pm= \pm \sqrt{-800}$ in
Eq.~(\ref{realt}).

The classification of the reality/imaginarity of the energies
proceeds as before. One finds four complex energy roots in the
interval of $z \in (1,81)$, the quadruplet of the purely imaginary
roots at the larger ${z} \in (81,\infty)$ and, finally, two real and
two imaginary energies at the smallest eligible $z \in (0,1)$. The
evaluation of the EP5 transition matrix
 $$
 Q^{(5)}=\left[ \begin {array}{ccccc} -3456&-576\,i&48&-4\,i&1
 \\\noalign{\medskip}-1728\,i&-144&-48\,i&8&0\\\noalign{\medskip}-1728
\,i\sqrt {6}&144\,\sqrt {6}&24\,i\sqrt
{6}&0&0\\\noalign{\medskip}1728
\,i&-432&0&0&0\\\noalign{\medskip}-3456&0&0&0&0\end {array} \right]
 $$
makes the whole construction completed.

\section{Discussion\label{disskuss}}

Once a Hamiltonian admits complex eigenvalues, we lose the
possibility of its Hermitization, i.e., of the unitarity of quantum
evolution based on the introduction of a suitable {\it ad hoc\,}
metric $\Theta=\Theta(H)$ amending the Hilbert space. Two consistent
physical interpretations of the model remain available. In one we
can treat the model as an incomplete, effective description of the
so called open quantum system (cf. \cite{Nimrod,Ingrid}). In an
alternative interpretation one returns to classical physics and one
accepts the loss of unitarity as a characteristic feature of the
evolution process in question.

In our present paper we considered both the real- and
complex-spectrum scenarios. We pointed out the similarities as well
as differences. A number of technical challenges was considered,
indicating that the most difficult part of the build-up of an
acceptable {\em quantum\,} non-Hermitian unitary model may be seen
in its necessary Hermitization, i.e., in the construction of the
Hamiltonian-dependent metric $\Theta(H)$. In comparison, the most
difficult part of the proposals of the analogous {\em
classical-physics\,} models relates to their realization in the
laboratory.

Let us now add a few remarks. We shall stress a few most important
differences, mathematical as well as phenomenological. We also
intend to emphasize the existence of the shared features including
(1) the phenomenological relevance of the EP singularities  and (2)
the practical feasibility of predictions.

%

\subsection{Physics represented by classical non-Hermitian Hamiltonians}

Inside the non-quantum setup the most difficult obstacles are
currently well known to emerge in a design of the experimental
realizations. The absence of the necessity of the construction of
the physical inner products simplifies mathematics. The majority of
the existing experiments seems connected with the most elementary
picture of non-Hermitian eigenvalue degeneracies {\it alias\,} phase
transitions (we restricted our consideration to the
finite-dimensional models mainly due to this reason).

The mathematical core of our message may be seen in the construction
of the models in which just a few eigenvalues merge at an elementary
exceptional point of order $K$ (EPK) with a not too large $K$. We
just returned to our constructive studies of the quantum models
possessing higher-order exceptional points \cite{catast}, and we
just omitted the requirement of the reality of the spectrum. Such a
change of perspective was inspired by the recent growth of interest
in sophisticated experiments localizing the exceptional points of
higher order in classical systems. We felt mostly
inspired by the recent conference report \cite{Cart} in which the
exceptional points of the third order (EP3) were studied in a system
of coupled waveguides. We also noticed that in Ref.~\cite{sensing},
optical microcavities were identified as prime candidates for the
sensing applications of the EPs. The third-order exceptional points
were identified, in Ref.~\cite{sensing2}, for a system of the three
coupled micro-rings made from a semiconductor material. Using the
same, rather restricted and more or less purely numerical
mathematics people also considered the waves coupled in acoustic
cavities with asymmetric losses in which the realization of the
fourth-order EP4 proved feasible \cite{PRX}.

We decided to concentrate on the not yet fully clarified theoretical
aspects of the similar experiments. In contrast to the difficulties
of the practical fine-tuning of parameters in experiments, the
construction of the non-quantum theoretical models supporting the
$K-$th-order exceptional points (EPK) may already be declared well
advanced at present. These developments proceeded along several
independent lines. People tested and, subsequently, widely accepted
that, first of all, one can hardly move beyond the $N=3$ matrix
models without the heuristically helpful requirement of ${\cal
PT}-$symmetry \cite{Nje4}.

The acceptance of ${\cal PT}-$symmetry opened the way towards the
not entirely expected non-numerical (albeit computer-assisted and
symbolic-manipulation-based \cite{Gerdt}) constructions of suitable
$H$s and, in particular, to the localization of their EPs via closed
formulae \cite{catast,maximal,tridiagonal}. Another efficient
EP-construction tool has been found in the standard self-adjoint toy
models known and used in condensed-matter physics. Via a
straightforward replacement of some of their real parameters by the
purely imaginary or complex quantities it was possible to preserve
the underlying algebraic solvability features. Interesting
generalizations were obtained for the ${\cal PT}-$symmetrized
Su-Schrieffer-Heeger model \cite{Ruzicka}, for the Aubry-Andr\'e and
Harper models \cite{Jogle}, etc.

In the ${\cal PT}-$symmetrized Bose-Hubbard model of Graefe et
al.~\cite{Uwe} the authors were particularly successful when they
recalled the representation theory of angular momentum algebra. This
facilitated their study of the unfolding of the EP degeneracy under
perturbations. In our present paper, another feature of the model
(viz, the matrix tridiagonality and complex symmetry of the
Hamiltonian) were found almost equally useful for generalizations,
especially at the lower dimensions $N$. Along these lines we
discovered the existence of a family of new, in general non-unitary
and non-Bose-Hubbard models possessing higher-order exceptional
points.

\subsection{Mathematics behind the exceptional-point unfolding}


The turn of our attention to classical optics enabled us to avoid
various mathematical challenges connected with the proper
probabilistic interpretation of the evolving state vectors
$|\psi\kt$ in the consequent quantum setting. We did not need to
consider the Hermitian-conjugate form of Eq.~(\ref{SEtd}) which must
be solved in the full-fledged non-Hermitian quantum mechanics
\cite{NIP}. We also did not need to insist on the reality of the
eigenvalues of $H$ itself which may be necessary for the very
observability of quantum systems. At the same time, our interest in
the systems near EPs may be perceived as shared by both the
classical and quantum physicists, in both cases being separated into
their experimental and theoretical subcategories.

One of the most interesting questions emerging in the vicinity of
EPs concerns the role of perturbations in a removal of the
EP-related spectral degeneracy. A key mathematical subtlety of such
an ``EP-unfolding'' reflects the fact that one has to distinguish
between the unfolding of a single EP of order $K$ and the unfolding
of a family of the lower-order EPs of the respective orders $K_1$,
\ldots, $K_n$ such that $K_1+\ldots +K_n=K$. In mathematical
language, one has to speak here about the so called ``cycles''
\cite{Kato}. For our present purposes, an optimal clarification of
physics behind such a scenario may be mediated by examples.

\subsubsection{The cycles and their degeneracy at $N=3$}

In the constructive analysis of the non-Hermitian
three-by-three-matrix toy model Hamiltonian of Ref. \cite{eva} the
authors emphasized that the existence of the standard, EP3-related
Jordan-block canonical form
 \be
 J^{(3)}(E)=\left [\begin {array}{ccc}
    E&1&0
 \\{}0&E&1
 \\{}0&0&E
 \end {array}\right ]\,
 \label{3hisset}
 \ee
of their three-dimensional tridiagonal toy model Hamiltonian implies
that in the vicinity of the EP3 singularity the eigenenergies may be
perceived as represented by the analytic function which lives on a
three-sheeted Riemann surface (see, e.g., pp. 63 - 65 in
\cite{Kato}) for details). This is a purely theoretical feature of
the model which is in a one-to-one correspondence with the
possibility of the experimental confirmation that whenever one
succeeds in circumscribing the singularity, three circles are needed
for the system to return to its initial state, i.e., in the language
of quantum mechanics, to the initial wave function. In parallel, the
authors of Ref. \cite{eva} also added a remark that after some other
choice of the parameters in their Hamiltonian, one can encounter an
alternative scenario in which the system returns to its initial
state after the mere {\em two\,} circles.

An explanation of the apparent paradox is easy: the $K=2$ nature of
the new situation will merely reflect the partial survival of the
diagonalizability of the Hamiltonian. The canonical form of $H$ will
be mediated by Eq.~(\ref{realt}) in which one obtains the $K=2$
Jordan-block result of the following form
 \be
 J^{(1+2)}(E',E)=\left [\begin {array}{ccc}
    E'&0&0
 \\{}0&E&1
 \\{}0&0&E
 \end {array}\right ]\,.
 \label{3hisset}
 \ee
Thus, a cycle of lower order appears here due to the accidental
degeneracy $E' \to E$ between a non-EP and an EP energy level.


\subsubsection{Degenerate cycles at $N=4$}

Similar effects may be encountered also at the higher matrix
dimensions $N$ of course. The explanation is given by the
relationship between the number of circles and the respective
dimensions $K_j$ of the canonical Jordan blocks given by the so
called periods of the cycles of the eigenvalues in the vicinity of
the degenerate EPs (cf., e.g., p. 65 and equation Nr. (1.6) in
\cite{Kato})).

The canonical form of our present $N=4$ complex symmetric
Hamiltonian (\ref{tripa}) can mimic the reduction phenomenon at the
EP2 parameters $B=-27$, $A=64$ and $z=81$, yielding, via
Eq.~(\ref{realt}), the mere $K=2$ Jordan-block canonical
representation
 \be
 J^{(1+1+2)}(E',E'',E)=\left [\begin {array}{cccc}
  E'&0&0&0
 \\
  0&  E''&0&0
 \\{}0& 0&E&1
 \\{}0& 0&0&E
 \end {array}\right ]\,,
 \ \ \ \ E=0\,.
 \label{42hisset}
 \ee
This means that, hypothetically, the system returns to its original
state after the mere two circles circumscribing the EP singularity.

Along the same lines one could even get an exhaustive classification
of the alternative scenarios. For the four-dimensional Hamiltonians,
for example, one could complement the above-mentioned EP4 and
single-EP2 scenarios by the single-EP3 possibility
 \be
 J^{(1+3)}(E',E)=\left [\begin {array}{cccc}
  E'&0&0&0
 \\
  0&  E&1&0
 \\{}0& 0&E&1
 \\{}0& 0&0&E
 \end {array}\right ]\,
 \label{43hisset}
 \ee
or, finally, by its double-EP2 alternative
 \be
 J^{(2+2)}(E',E)=\left [\begin {array}{cccc}
  E'&1&0&0
 \\
  0&  E'&0&0
 \\{}0& 0&E&1
 \\{}0& 0&0&E
 \end {array}\right ]\,.
 \label{43hisset}
 \ee
An exhaustive classification of the five-dimensional (or higher)
non-equivalent scenarios would be slightly more complicated but
equally straightforward.

\subsection{The physics behind the {\em quantum\,} non-Hermitian models}

In quantum systems one has to deal with several paradoxical new
aspects of the old question of the stability or instability of
systems exposed to small perturbations. In the quantum
unitary-evolution setting the emergence of instabilities {\it
alias\,} quantum catastrophes is not always sufficiently well
understood and explained in the literature, in spite of being one of
the most characteristic consequences of the occurrence of EPs. Along
these lines, a complementary inspiration of our research was
provided also by the conventional Hermitian treatment of quantum
information. The ${\cal PT}-$symmetrization recipe opened the way,
e.g., towards the perfect-transfer-of-states protocol which has been
based on the choice and treatment of the Hamiltonian as an angular
momentum in an external magnetic field \cite{Song}.

There exist two remarkable byproducts of the latter choices of $H$.
One of them lies in the exact, non-numerical description of the
critical behavior at the EP2 singularities even after a fairly
nontrivial hypercube-graph generalization of the model. This might
prove inspiring in the future. Another source of inspiration may be
found in Ref.~\cite{Teimo} where the authors developed a recursive
bosonic quantization technique which is able to generate the
generalized classes of the ${\cal PT}-$symmetric networks and other
classical photonic structures exhibiting numerous interesting
topological and symmetry features (cf., e.g., Figure Nr.~2 in {\it
loc.~$\!$cit.}).

Once one turns attention to non-Hermitian quantum systems, one
reveals the existence of a number of paradoxes created, often, by
the lack of the necessary mathematical insight. The physics of
quantum systems represented by non-Hermitian phenomenological
observables $H^{(N)}$, $\Lambda^{(N)}$, \ldots (with real spectra
and with any matrix dimension $N$, finite or infinite) becomes
particulary interesting in the vicinity of their EP-singularity
boundaries. For illustration, a few quantitative studies of the
emergence of related quantum phase transitions may be found in
\cite{catast}. The authors of some other studies did not always keep
in mind the fundamental requirements of the consistent probabilistic
interpretation of their quantum models. This may lead to
misunderstandings, indeed.

%

\subsubsection{The strength-of-perturbation
paradox\label{precedingp}}





The physics represented by non-Hermitian operators is still under
intensive development. In most cases, one just has to avoid certain
more or less elementary misunderstandings. Many of them were already
clarified in review \cite{ali}. One encounters their subtler forms
also in the more recent literature. For example, in
Ref.~\cite{Viola} the authors claimed to have detected ``unexpected
wild properties of operators familiar from PT-symmetric quantum
mechanics'' and, as a consequence, they ``propose giving the
mathematical concept of the pseudospectrum a central role in quantum
mechanics with non-Hermitian operators.'' The phenomenologically
disturbing ``immanent instability'' claims are also illustrated, by
the authors of Ref.~\cite{Viola}, via a number of $N=\infty$
ordinary differential models.

Despite the use of a high-quality functional analysis the authors'
correct mathematical results are accompanied by their misleading
physical interpretation. We cannot endorse their claims. The point
lies very close to our above discussion: One must distinguish
between the classical, Maxwell-equation systems and the unitary
quantum models. Exclusively in such a case the picture of the
``wild'' physics is realistic (cf. also a number of further examples
of the non-Hermiticity-caused instabilities in \cite{Trefethen}). In
the quantum-physics approach, in contrast, it would be necessary to
amend the inner-product metric $I \to \Theta$ making the quantum
version of the theory consistent by means of the necessary
transition from the auxiliary, ``user-friendly'' but ``false''
Hilbert space ${\cal H}^{(F)}$ to its correct, physical,
``standard'' alternative ${\cal H}^{(S)}$.

The latter link is missing in \cite{Viola}. In more detail, the
purely mathematical explanation of the misunderstanding lies in the
use of the ``false'' pseudospectrum
$$
 \Lambda_\epsilon^{(F)}(A)=\{
 \lambda \in \mathbb{C}\ |\
 \exists \psi \in {\cal H}^{(F)} \setminus \{0\}, \exists V : (A+V)\psi =
 \lambda \psi, ||V||_{_{{\cal H}^{(F)}}} \leq \epsilon
 \}\,.
 $$
This definition is inappropriate, based on the use of the norm of
the perturbation $V$ in the {\em manifestly unphysical\,} Hilbert
space ${\cal H}^{(F)}$. The analysis of the latter, ill-defined (or,
better, unphysical) pseudospectrum is, in the unitary quantum-theory
setting, irrelevant. The calculation does not take into account the
realistic metric, i.e., the different geometry of the unique
physical Hilbert space ${\cal H}^{(S)}$.

In quantum world the influence of the perturbations must necessarily
be characterized by another, correctly defined, {\em
metric-dependent\,} pseudospectrum
 $$
 \Lambda_\epsilon^{(S)}(A)=\{
 \lambda \in \mathbb{C}\ |\
 \exists \psi \in {\cal H}^{(S)} \setminus \{0\}, \exists V : (A+V)\psi =
 \lambda \psi, ||V||_{_{{\cal H}^{(S)}}} \leq \epsilon
 \}\,.
 $$
In spite of the technically much more complicated nature of the
latter, amended pseudospectrum, one cannot expect that its
evaluation would lead to the ``unexpected wild properties'' of any
admissible ${\cal PT}-$symmetric operators \cite{admis}.
Indeed, whenever one satisfies the (necessary) hidden-Hermiticity
constraint (\ref{requi}), and whenever the perturbations remain
small in the correct physical geometry of Hilbert space ${\cal
H}^{(S)}$, one {\em does not\,} encounter any instabilities.

%
%


\subsubsection{The Bose-Hubbard unfolding paradox\label{papapa}}

%

In the non-Hermitian but ${\cal PT}-$symmetric Bose-Hubbard quantum
model the authors of Ref.~\cite{Uwe} studied the phenomenon of the
unfolding (i.e., of the removal of degeneracy) of the higher-order
exceptional-point energy under a well-defined perturbation $V$.
Unfortunately, they also did not formulate the task in a consistent
manner, i.e., in the correct physical Hilbert space ${\cal H}^{(S)}$
with metric $\Theta^{(correct)}(H_0+V)$. For this reason, the
description of the unfolding of the multiply degenerate (i.e.,
higher-order EP) Bose-Hubbard bound-state energies
$E_n(\gamma^{(EP)})$ after perturbation \{cf. sections \# 4
(numerical results) and \# 5 (perturbation results) in
Ref.~\cite{Uwe}\} must be characterized as incomplete. The main
reason is that the self-consistent nature of perturbation theory in
non-Hermitian quantum picture makes the calculations, in effect,
non-linear. We have to keep in mind that
 $$
 \Theta_0=\Theta(H_0) \neq
 \Theta^{(correct)}(H_0+V)=\Theta_0+K(\Theta_0,H_0,V)\,.
 $$
For each separate perturbation $V$ we have to repeat the solution of
Eq.~(\ref{requi}). For the ``sufficiently small'' perturbations
(whatever it means) we can simplify the process and neglect the
higher-order terms. Equation~(\ref{requi}) gets replaced by
 \be
 H_0^\dagger\, K-K \,H_0= \Theta_0\,V-V^\dagger\, \Theta_0 \,
 \ee
i.e., by the implicit definition of the leading-order form of $K$.
Whenever one skips this apparently merely technical step, the
results cannot be declared physical.

In the ${\cal PT}-$symmetric Bose-Hubbard case the criticism of
subsection \ref{precedingp} re-applies. The proper probabilistic
interpretation cannot be provided without the knowledge of
$\Theta^{(correct)}(H_0+V)$. Without this knowledge, the perturbed
system is merely tractable as classical. Moreover, in \cite{Uwe},
the perturbed eigenvalues cease to be real and form the rings in
complex plane. The perturbed Bose-Hubbard quantum system ceases to
be observable so that it must be declared non-unitary  and unstable.


Beyond the concrete Bose-Hubbard model the generic change of
non-Hermitian perturbations will always imply a nontrivial change of
the metric $\Theta$. The mutual relationship between these changes
can be best studied in quantum systems with small $N \ll \infty$.
The correlations become particularly relevant near the dynamical
EP-singularity. In the EP limit the eligible
Hamiltonian-Hermitization metrics $\Theta(H)$ will cease to exist
because all of the candidates for the metric (i.e., {\em all\,} of
the solutions of Eq.~(\ref{requi})) will cease to be invertible. The
geometry of the physical Hilbert space will become, in the EP
vicinity, strongly anisotropic \cite{lotor}.


\section{Summary\label{sedruha7}}

Although the notion of exceptional point (EP) emerged, in the
context of perturbation theory, in mathematics \cite{Kato}, it very
quickly acquired applications in several branches of physics
\cite{Heiss}. The conference ``The Physics of Exceptional Points''
in 2010 \cite{Stellenbosch} covered, for example, the domains of
physics as different as the study of Bose-Einstein condensates and
of the light-matter interactions, or the behavior of molecules
during photo-dissociation and the phase transitions related to the
spontaneous breakdown of ${\cal PT}-$symmetry, or the questions of
stability in many-body quantum systems as well as in classical
magnetohydrodynamics. Still, the most immediate motivation of the
meeting seems to have been provided by the series of speculations
and experiments \cite{experdva} which demonstrated the presence of
an EP singularity in the eigenvalue and eigenvector spectra of
various classical devices. In \cite{nemci}, for example, two modes
of a certain classical electromagnetic microwave billiard (i.e.,
eigenvectors $\psi_1(z)$ and $\psi_2(z)$ of its ``Hamiltonian''
$H(z)$) were shown to coalesce,  at the EP singularity (i.e., at a
parameter $z=z^{(EP)}$), with a phase difference of $\pi/2\,$.

The EP singularity in question was shown to be of the square-root
type (abbreviated as EP2). From the point of view of elementary
linear algebra this means that the two-by-two matrix $H(z^{(EP)})$
ceased to be diagonalizable and that it acquired a canonical form of
two-by-two Jordan block. In the language of functional analysis the
related two eigenvalues $E_1(z)$ and $E_2(z)$ of $H(z)$ with $z \neq
z^{(EP2)}$ could be called ``cyclic'', of period two (cf. p. 64 in
\cite{Kato}). The importance of this feature in physics can be
deduced from the very logo of conference \cite{Stellenbosch}. In an
illustration of the consequences of the cyclicity this logo samples
the intensity of the electromagnetic field in the microwave during a
step-by-step variation of parameter $z=z(t)$. The parameter is made
to circumscribe its critical value $z^{(EP2)}$ so that one can see
that the billiard can only return to its initial state after {\em
two\,} circles.


In our present paper we emphasized that under the assumption of
complex symmetry and tridiagonality of Hamiltonians, the transition
to the more general $N$ by $N$ matrices $H^{(N)}(z^{(EP)})$ can been
found, both theoretically and experimentally, feasible.
%
%
A decisive encouragement can be sought in the purely empirical fact
that in the dedicated literature, virtually all of the successful
benchmark Hamiltonians $H^{(N)}(z)$ are being chosen, even at the
lowest dimensions $N=2$ and $N=3$, in the very specific complex
symmetric and tridiagonal matrix forms exhibiting a number of
features shared with the angular-momentum representation methods,
say, of Refs.~\cite{Uwe,maximal} or \cite{Teimo}. In this spirit we
performed here an extended search using a straightforward linear
algebraic method. Considering just a few-parametric class of the
eligible candidates $H^{(N)}(z)$ we filled the gap, on the side of
theory, up to $N=4$ and $N=5$. The resulting models appeared,
unexpectedly, exactly solvable. We believe that the experimental
simulations based on these matrices will prove equally user-friendly
in the future, especially in the vicinity of the various EP-limiting
cases.

\section*{Acknowledgement}

Work supported by GA\v{C}R Grant Nr. 16-22945S.



\end{document}